\documentclass[aps,twocolumn,superscriptaddress,prd]{revtex4-1}

\usepackage{bm,graphicx,textcomp,amssymb,amsmath,dcolumn,color}

\usepackage{slashed}

\usepackage{tabularx}
\usepackage{ctable}
  
\usepackage[utf8]{inputenc}
\usepackage[T1]{fontenc}

\newcommand{\nn}{\nonumber\\}

\newcommand{\bea}{\begin{eqnarray}}
\newcommand{\ea}{\end{eqnarray}}
\newcommand{\eea}{\end{eqnarray}}
\newcommand{\ord}{\,{\cal O}}

\begin{document}

\title{Detection schemes for quantum vacuum diffraction and birefringence}

\author{N.~Ahmadiniaz} 
\affiliation{Helmholtz-Zentrum Dresden-Rossendorf, Bautzner Landstra\ss e 400, 01328 Dresden, Germany} 

\author{T.E.~Cowan}
\affiliation{Helmholtz-Zentrum Dresden-Rossendorf, Bautzner Landstra\ss e 400, 01328 Dresden, Germany} 
\affiliation{Institut f\"ur Kern-und Teilchenphysik, Technische Universit\"at Dresden, 01062 Dresden, Germany}

\author{J.~Grenzer}
\affiliation{Helmholtz-Zentrum Dresden-Rossendorf, Bautzner Landstra\ss e 400, 01328 Dresden, Germany}

\author{S.~Franchino-Vi\~nas}
\affiliation{Helmholtz-Zentrum Dresden-Rossendorf, Bautzner Landstra\ss e 400, 01328 Dresden, Germany} 

\author{A.~Laso~Garcia}
\affiliation{Helmholtz-Zentrum Dresden-Rossendorf, Bautzner Landstra\ss e 400, 01328 Dresden, Germany}

\author{M.~\v{S}m\'id}
\affiliation{Helmholtz-Zentrum Dresden-Rossendorf, Bautzner Landstra\ss e 400, 01328 Dresden, Germany}

\author{T.~Toncian}
\affiliation{Helmholtz-Zentrum Dresden-Rossendorf, Bautzner Landstra\ss e 400, 01328 Dresden, Germany}

\author{M.A.~Trejo}
\affiliation{Helmholtz-Zentrum Dresden-Rossendorf, Bautzner Landstra\ss e 400, 01328 Dresden, Germany} 

\author{R.~Sch\"utzhold}
\affiliation{Helmholtz-Zentrum Dresden-Rossendorf, Bautzner Landstra\ss e 400, 01328 Dresden, Germany} 
\affiliation{Institut f\"ur Theoretische Physik, Technische Universit\"at Dresden, 01062 Dresden, Germany}

\begin{abstract}
Motivated by recent experimental initiatives, such as at the 
Helmholtz International Beamline for Extreme Fields (HIBEF) 
at the European X-ray Free Electron Laser (XFEL), we calculate 
the birefringent scattering of x-rays at the combined field of 
two optical (or near-optical) lasers and compare various scenarios. 
In order to facilitate an experimental detection of quantum vacuum diffraction and 
birefringence, special emphasis is placed on scenarios where the difference between the initial and final x-ray photons is maximized.
Apart from their polarization, these signal and background photons may 
differ in propagation direction (corresponding to scattering angles in the mrad regime) 
and possibly energy.
\end{abstract}

\date{\today}

\maketitle

\section{Introduction}

The quantum vacuum is not just empty space -- as the ground state of interacting quantum field theories,
it displays a complex structure, entailing many fascinating phenomena.
For example, quantum electrodynamics (QED) predicts that the quantum vacuum should behave like a non-linear 
medium and thus display effects such as diffraction and refraction as well as birefringence under the influence 
of a strong electromagnetic field 
\cite{euler-35,euler+heisenberg,sauter-31,schwinger-51,karplus+neuman}. 

Related phenomena have already been observed in the form of Delbr\"uck scattering of $\gamma$ rays on the 
Coulomb fields of nuclei, which can be interpreted as quantum vacuum refraction 
\cite{meitner-33,bethe+rohrlich,costantini-71,DESY-exp,milshtein-1,milshtein-2,akhm-98,rev1,rev2,rev3,rev4}, 
or
the interaction of the Coulomb fields of two nuclei almost colliding with each other at 
ultra-high energies and the resulting emission of a pair of $\gamma$ quanta 
\cite{atlas,denter-13}.

In contrast, we are mostly interested in scenarios without additional massive particles 
(such as nuclei, see also \cite{coulombass}) and at sub-critical scales in the following: 
First, the relevant energies and momenta should be well below the electron mass $m\approx0.51~{\rm MeV}/c^2$ 
and thus the characteristic length and time scales well above the reduced Compton length 
$\lambdabar=\hbar/(mc)\approx386~\rm fm$.  
Second, the involved field strengths should be well below the critical 
fields of QED, i.e., $E_{\rm crit}=m^2c^3/(\hbar q)\approx 1.3\times10^{18}\,{\rm V}/{\rm m}$ as well as  
$B_{\rm crit}=E_{\rm crit}/c\approx4.4\times10^9~\rm T$.

Many theoretical and several experimental investigation have been devoted to this sub-critical regime. 
For polarizing the quantum vacuum, one could use a strong and quasi-static magnetic field 
\cite{della-16,zava-12,pvlas,zava-08,zava-06,PVLAS-20,OVAL,BMV,hartman-17,battesti-18}
or the focus of an optical or near-optical laser or XFEL as the pump field, see, e.g., 
\cite{heinzl-06,di-piazza-06,inada-14,yamaji-16,Schlenvoigt-2016,Karbstein-2016,king-18,
lund-06,inada-2017,dipiazza-07,tommasini-09,
tommasini-10,king-12,gies1,gies2,gies3,gies-09,gies-15,grote-15,ahm-20,gies-22,gies-XFEL-XFEL}. 
For the detection of the induced vacuum non-linearity, one could also employ an optical or near-optical 
laser or an XFEL as the probe field, see, e.g., 
\cite{heinzl-06,di-piazza-06,inada-14,yamaji-16,Schlenvoigt-2016,Karbstein-2016,lund-06,inada-2017,ahm-20,
coulombass}. 
Motivated by experimental facilities such as HIBEF and the quest to maximize the signal,
we consider an XFEL probe (whose shorter wavelength yields a larger interaction probability in a given volume) 
and an optical pump field (which facilitates a high intensity). 

In spite of the efforts so far, neither quantum vacuum birefringence nor quantum vacuum diffraction 
-- or, more generally, quantum vacuum non-linearity in the sub-critical regime -- have been conclusively 
verified in a laboratory experiment yet \cite{PVLAS-20,OVAL,BMV}. 
Apart from a pure verification of this fundamental QED prediction, such an experiment would also allow 
us to search for new phenomena beyond the standard model of particle physics because they could manifest 
themselves in measurable deviations from the QED predictions. 

In order to facilitate such an experiment, it is crucial to distinguish the signal 
(consisting of one or a few x-ray photons) from the background, i.e., the XFEL pulse. 
Since such a distinction purely based on the photon polarizations can be quite challenging,
it has been suggested to consider scenarios where the initial (background) and final (signal)
x-ray photons also differ in propagation direction and possibly energy, see, e.g.,  
\cite{karb-22,king-18}. 
Developing these ideas further, we propose and study scenarios (see Fig.~\ref{figure})
which maximize this difference, especially the momentum transfer and thus the scattering angle.

\section{Effective Lagrangian}

Since all the involved scales are supposed to be sub-critical, we start from the generalized lowest-order 
Euler-Heisenberg Lagrangian ($\hbar=c=\epsilon_0=\mu_0=1$)
\bea
\label{EHL}
\mathfrak{L}
=
\frac{1}{2}\left(\mathfrak{E}^2-\mathfrak{B}^2\right)+
a\left(\mathfrak{E}^2-\mathfrak{B}^2\right)^2
+b\left(\mathfrak{E}\cdot\mathfrak{B}\right)^2
\,,
\ea
in terms of the total electric $\mathfrak{E}$ and magnetic $\mathfrak{B}$ fields. 
In QED, the two parameters $a$ and $b$ are given by $b=7a$ and $a=q^4/(360\pi^2m^4)$, 
for reviews see 
\cite{review-king+heinzl-2016,review-karbstein-2020,review-dipiazza-12,review-dunne-12,Toll-PhD}. 
However, in order to accommodate possible deviations from the standard model, 
we shall keep them as general. 
For example, a coupling to an axion field would typically manifest itself in a modification 
of the $b$ parameter while $a$ remains unchanged, see, e.g., \cite{siki-07,siki-84,raffel-88,zyla-20}.

The propagation of the probe field ${\bf E}$ and ${\bf B}$
(i.e., the XFEL) in the presence of a given pump field ${\bf E}_{\rm L}$ and ${\bf B}_{\rm L}$
(i.e., the optical laser) can be studied by inserting 
the split 
$\mathfrak{E}={\bf E}_{\rm L}+{\bf E}$ and $\mathfrak{B}={\bf B}_{\rm L}+{\bf B}$
and linearizing the equations of motion, leading to the effective Lagrangian  
\bea
\label{probe}
\mathfrak{L}_{\rm eff}
=
\frac12\left[
{\bf E}\cdot({\bf 1}+\delta\epsilon)\cdot{\bf E}-
{\bf B}\cdot({\bf 1}-\delta\mu)\cdot{\bf B}
\right] 
+{\bf E}\cdot\delta\Psi\cdot{\bf B}
\,,
\nn
\ea
with the symmetric permittivity/permeability tensors 
\bea
\delta\epsilon^{ij}
&=&
8aE_{\rm L}^i E_{\rm L}^j 
+2bB_{\rm L}^i B_{\rm L}^j 
+4a\delta^{ij}({\bf E}_{\rm L}^2-{\bf B}_{\rm L}^2)
\,,
\nn
\delta\mu^{ij}
&=&
2bE_{\rm L}^i E_{\rm L}^j 
+8aB_{\rm L}^i B_{\rm L}^j 
-4a\delta^{ij}({\bf E}_{\rm L}^2-{\bf B}_{\rm L}^2)
\,,
\ea
plus the symmetry-breaking contribution
\bea
\label{symmetry-breaking}
\delta\Psi^{ij}
&=&
-8aE_{\rm L}^i B_{\rm L}^j 
+2bB_{\rm L}^i E_{\rm L}^j 
+2b\delta^{ij}({\bf E}_{\rm L}\cdot{\bf B}_{\rm L})
\,,
\ea
which describe the polarizability of the QED vacuum. 
Note that the latter tensor is not symmetric 
$\delta\Psi^{ij}\neq\delta\Psi^{ji}$. 

As is well known, the linearized equations of motion generated by~\eqref{probe} can be cast into the 
same form as the macroscopic Maxwell equations in a medium 
$\nabla\cdot{\bf D}=0$, $\nabla\cdot{\bf B}=0$, 
$\nabla\times{\bf E}=-\partial_t{\bf B}$, and 
$\nabla\times{\bf H}=\partial_t{\bf D}$, provided that we introduce the 
electric 
${\bf D}=({\bf 1}+\delta\epsilon)\cdot{\bf E}+\delta\Psi\cdot{\bf B}$ 
and magnetic 
${\bf H}=({\bf 1}-\delta\mu)\cdot{\bf B}-\delta\Psi^{\rm T}\cdot{\bf E}$ displacement fields. 

\section{Scattering Theory}

The scattering of the x-ray photons can be calculated via various options, 
e.g., time-dependent perturbation theory of quantum fields or the  
photon emission picture (see, e.g., \cite{gies3}). 
In the following, we shall employ classical scattering theory \cite{jackson,coulombass,di-piazza-06}, 
but adapted to the case of oscillating contributions in $\delta\epsilon$, $\delta\mu$ and $\delta\Psi$.
To this end, we combine the above Maxwell equations to 
\bea
\label{box}
\Box{\bf D}
=
\nabla\times\left[\nabla\times({\bf D}-{\bf E})\right]
+\partial_t\left[\nabla\times({\bf H}-{\bf B})\right]
={\bf J}^{\rm eff}
\quad
\ea
where the effective source term $ {\bf J}^{\rm eff}$ 
on the right-hand side encodes the quantum vacuum non-linearity. 
Since this term is very small, we may employ the usual Born approximation.
Thus, we split the XFEL field ${\bf D}$ into an ingoing plane wave 
${\bf D}^{\rm in}$ with $\omega_{\rm in}$ and ${\bf k}_{\rm in}$ 
plus a small scattering contribution ${\bf D}^{\rm out}$ 
induced by vacuum polarizability $\delta\epsilon$, $\delta\mu$, and $\delta\Psi$.

These quantities $\delta\epsilon$, $\delta\mu$, and $\delta\Psi$ depend on the 
optical laser (i.e., pump) fields ${\bf E}_{\rm L}$ and ${\bf B}_{\rm L}$ and thus 
on time. 
Here we assume that this pump field is generated by the superposition of two optical lasers,
i.e., ${\bf E}_{\rm L}={\bf E}_1+{\bf E}_2$ and ${\bf B}_{\rm L}={\bf B}_1+{\bf B}_2$
which oscillate with frequencies $\omega_1$ and $\omega_2$, respectively. 
Due to the resulting oscillatory time-dependence of $\delta\epsilon$, $\delta\mu$, and $\delta\Psi$,  
the outgoing field ${\bf D}^{\rm out}$ contains various frequency contributions
$\omega_{\rm out}=\omega_{\rm in}\pm\omega_1\pm\omega_2$ (similar to Floquet bands).
Since the combinations $\omega_{\rm out}=\omega_{\rm in}+\omega_1+\omega_2$ and 
$\omega_{\rm out}=\omega_{\rm in}-\omega_1-\omega_2$ are typically not allowed by 
energy-momentum conservation (see below), we focus on 
$\omega_{\rm out}=\omega_{\rm in}+\omega_1-\omega_2$ 
and  $\omega_{\rm out}=\omega_{\rm in}-\omega_1+\omega_2$ in the following.

Then, Eq.~\eqref{box} turns into a Helmholtz equation 
\bea
\Box{\bf D}_{\omega}^{\rm out}
&=&
-\left(\nabla^2+\omega^2_{\rm out}\right){\bf D}_\omega^{\rm out}
=
{\bf J}_\omega^{\rm eff}
\,,
\ea
which can be solved by the usual Greens function. 
In the far field, we thus obtain the scattering amplitude 
\bea
\label{amplitude}
{\mathfrak A}=\frac{1}{4\pi|{\bf D}_\omega^{\rm in}|}\,{\bf e}_{\rm out}\cdot
\int d^3r\,\exp\{-i{\bf k}_{\rm out}\cdot{\bf r}\}\,{\bf J}_\omega^{\rm eff}
\,,
\ea
depending on the 
momentum ${\bf k}_{\rm out}$ and polarization ${\bf e}_{\rm out}$
of the outgoing x-ray photon. 
Since $\omega_1$ and $\omega_2$ are optical or near optical frequencies of order $\ord(\rm eV)$ 
while $\omega_{\rm out}$ and $\omega_{\rm in}$ are x-ray frequencies on the keV regime, 
we may neglect small terms such as $\omega_1/\omega_{\rm out}$ 
(in comparison to a non-vanishing leading-order term) 
in the following and approximate $\omega_{\rm out}\approx\omega_{\rm in}$. 

Furthermore, the integral in Eq.~\eqref{amplitude} simplifies drastically if we approximate the 
two optical (pump) lasers by plane waves with momenta ${\bf k}_1$ and ${\bf k}_2$ 
and polarizations ${\bf e}_1$ and ${\bf e}_2$. 
In this case, the $d^3r$ integral just corresponds to momentum conservation and the 
scattering amplitude simplifies to 
\bea
\label{delta}
{\mathfrak A}
\approx
8\pi^2
{\mathfrak F}
E_1E_2
\omega_{\rm out}^2
\delta^3({\bf k}_{\rm out}-{\bf k}_{\rm in}-{\bf k}_1+{\bf k}_2)
\ea
where  
${\mathfrak F}
({\bf e}_1,{\bf e}_2,{\bf e}_{\rm in},{\bf e}_{\rm out},{\bf n}_1,{\bf n}_2,{\bf n}_{\rm in},{\bf n}_{\rm out})$
is a purely algebraic expression of the four polarization vectors ${\bf e}_I$ and propagation direction unit 
vectors ${\bf n}_I={\bf k}_I/\omega_I$. 

In order to obtain a compact expression for $\mathfrak{F}$, 
we will make explicit use of its symmetry under permutations of the indices
in the kinematical variables; 
we will thus write a single contribution and permutations thereof should be added.  
Using this convention,  $\mathfrak{F}$ is given as
\bea
\mathfrak{F}=a\mathfrak{F}_a+b\mathfrak{F}_b,
\ea
where we have defined 
\bea
\label{general-F}
\mathfrak{F}_a
&=&
\Big[({\bf e}_2\cdot{\bf e}_{\rm in})-({\bf e}_2\times{\bf n}_2)\cdot({\bf e}_{\rm in}\times{\bf n}_{\rm in})\Big]\nonumber\\
&&
\Big[
\big({\bf e}_{\rm out}\times{\bf n}_{\rm out}\big)\cdot\big({\bf e}_1\times[{\bf n}_{\rm out}-{\bf n}_1]\big)
\Big]
%
%
\nonumber\\
&&+{\rm permutations\{1,2,in\}}
\nonumber,
\\
\mathfrak{F}_b
&=&
\frac{1}{2}\,\Big[\big({\bf n}_{\rm out}\times({\bf e}_1\times{\bf n}_1)-{\bf e}_1\big)\times{\bf n}_{\rm out}\Big]\cdot{\bf e}_{\rm out}
\nonumber\\
&&
\left[
{\bf e}_2\cdot({\bf e}_{\rm in}\times{\bf n}_{\rm in})
\right] 
+{\rm permutations\{1,2,in\}}
.
\ea
As explained above, we have to sum over permutations, i.e., for each expression 
$\mathfrak{F}_{a,b}\{\rm 1,2,in\}$, we have to add the other five combinations 
$\mathfrak{F}_{a,b}\{\rm 1,in,2\}$,
$\mathfrak{F}_{a,b}\{\rm 2,in,1\}$,
$\mathfrak{F}_{a,b}\{\rm 2,1,in\}$,
$\mathfrak{F}_{a,b}\{\rm in,1,2\}$, and 
$\mathfrak{F}_{a,b}\{\rm in,2,1\}$.


\section{Counter-propagating Case}\label{Counter-propagating Case}

Let us first consider the set-up already discussed in the literature \cite{heinzl-06,Schlenvoigt-2016}, 
where the XFEL interacts with a single counter-propagating optical laser, see Fig.~\ref{figure}a.
Thus we set $\omega_1=\omega_2$ and ${\bf e}_1={\bf e}_2$ as well as 
${\bf n}_1={\bf n}_2$ which implies $\omega_{\rm out}=\omega_{\rm in}$ 
and ${\bf n}_{\rm in}={\bf n}_{\rm out}$. 
%
Hence
we have ${\bf n}_{\rm in}={\bf n}_{\rm out}=-{\bf n}_1=-{\bf n}_2$ and 
${\mathfrak F}$ simplifies to 
\bea
\label{counter-propagating}
{\mathfrak F}
&=&
16a({\bf e}_{\rm in}\cdot{\bf e}_1)({\bf e}_{\rm out}\cdot{\bf e}_1)
\nn
&&
+4b{\bf e}_{\rm in}\cdot({\bf n}_{\rm in}\times {\bf e}_{\rm 1}) \, 
{\bf e}_{\rm out}\cdot ({\bf n}_{\rm in}\times {\bf e}_{1})
\,.
\ea
In this case, the energy-momentum transfer is exactly zero, i.e., the scattering angle vanishes. 
Thus the signal and background photons can only be distinguished by their polarization, 
which is an experimental challenge. 
A finite (albeit small) scattering angle can be induced by the spatial inhomogeneity of laser focus
(i.e., going beyond the plane-wave approximation), in close analogy to a lens (quantum vacuum refraction), 
see also
\cite{gies1,gies-15}.

\section{Crossed Beam Case}\label{Crossed Beam Case}

The above difficulty can be avoided by considering
a modified scenario where the pump field is generated by 
two optical laser beams with the same frequency $\omega_1=\omega_2$ and field strength, 
but different propagation directions ${\bf n}_1\neq{\bf n}_2$. 
In Ref.~\cite{king-18}, a fully perpendicular set-up with ${\bf n}_1\perp{\bf n}_2$ has been considered. 
However, in order to maximize the momentum transfer (i.e., have a large scattering angle), we propose 
the head-on collision of two optical laser beams where ${\bf n}_1=-{\bf n}_2$.  
Since the XFEL beam mainly jitters in horizontal direction, this geometry has the advantage 
of being quite robust if the optical lasers are also oriented in horizontal direction. 

Here, we consider the case ${\bf e}_1={\bf e}_2$ which yields the maximum electric field, 
%
%
but other polarizations (e.g., maximum magnetic field) would work as well.
Since $\omega_1=\omega_2$, 
the energy transfer vanishes again $\omega_{\rm out}=\omega_{\rm in}$, but we get a finite 
momentum transfer $\Delta{\bf k}=2\omega_1{\bf n}_1$. 
In order to satisfy $\omega_{\rm out}=\omega_{\rm in}$ and to transform this momentum transfer into a 
maximum 
scattering angle (in the mrad regime), we assume sending in the probe beam at a perpendicular direction 
${\bf n}_{\rm in}\perp{\bf n}_1$ and ${\bf n}_{\rm in}\perp{\bf e}_1$, see Fig.~\ref{figure}b.
In this case, we find 
\bea
\label{crossed}
{\mathfrak F}
=
4a({\bf e}_{\rm in}\cdot{\bf e}_1)({\bf e}_{\rm out}\cdot{\bf e}_1)
+
b({\bf e}_{\rm in}\cdot{\bf n}_1)({\bf e}_{\rm out}\cdot{\bf n}_1)
\,.
\ea
The polarization conserving signals at ${\bf e}_{\rm in}={\bf e}_{\rm out}={\bf e}_1$
and ${\bf e}_{\rm in}={\bf e}_{\rm out}={\bf n}_1$ would allow us to detect the 
parameters $a$ and $b$ separately, while we also get birefringent scattering 
${\bf e}_{\rm in}\neq{\bf e}_{\rm out}$ in the other directions. 

Note that the polarization conserving signals can also 
be distinguished from the background
by the momentum transfer yielding a scattering angle in the mrad regime.  
This scattering angle can be explained in close analogy to diffraction 
(as in ordinary Bragg scattering).
The spatial modulation (in ${\bf n}_1$ direction) acts like a grating which generates 
a Bragg peak at $\Delta{\bf k}=2\omega_1{\bf n}_1$.

\subsection{Cross Section}\label{Cross Section}

The amplitude~\eqref{amplitude} yields the differential cross section $d\sigma/d\Omega=|{\mathfrak A}|^2$. 
Obviously, inserting the plane-wave result~\eqref{delta}, this quantity would diverge due to the infinite 
volume. 
Thus, let us consider the more realistic case of a finite-size laser focus. 
To estimate the cross section, we rewrite the amplitude~\eqref{amplitude} by factoring out 
frequency, wavenumber and amplitude of the XFEL 
\bea
{\mathfrak A}=\frac{\omega^2_{\rm out}}{4\pi}
\int d^3r\,\exp\{i{\bf\Delta k}\cdot{\bf r}\}\,j^{\rm eff}
\,,
\ea
where $\Delta{\bf k}={\bf k}_{\rm in}-{\bf k}_{\rm out}$ is the momentum transfer.  
The renormalized source term $j^{\rm eff}({\bf r})$ now only depends on the optical laser and the 
XFEL polarization vectors.  
For example, for the case ${\bf e}_{\rm in}={\bf e}_{\rm out}={\bf e}_1$ discussed above, 
we find $j^{\rm eff}({\bf r})=8a{\bf E}_{\rm L}^2({\bf r})=8a[{\bf E}_1({\bf r})+{\bf E}_2({\bf r})]^2$.

Now let us assume that the spatial dependence of $j^{\rm eff}({\bf r})$ along the XFEL beam direction 
can be approximated (at least well inside the Rayleigh length)
by a simple (e.g., Gaussian) envelope function $f(r_\|)$ such that 
$j^{\rm eff}({\bf r})=f(r_\|)j^{\rm eff}_\perp({\bf r}_\perp)$.   
Furthermore, in view of energy conservation and $\omega_{\rm in}=\omega_{\rm out}\gg\omega_1=\omega_2$, 
the momentum transfer $\Delta{\bf k}$ is approximately perpendicular to ${\bf k}_{\rm in}$.  
Thus, we may approximate the integral over the solid angle $\int d\Omega$ 
by an integration over the transversal momentum transfer 
$\int d\Omega\approx\omega^{-2}_{\rm out}\int d^2{\bf\Delta k}_\perp$. 
This allows us to approximate the total cross section via 
\bea
\label{total}
\sigma\approx
\frac{\omega^2_{\rm out}L_\|^2}{4}
\int d^2r_\perp\left|j^{\rm eff}_\perp({\bf r}_\perp)\right|^2
\,,
\ea
with the interaction length $L_\|=\int dr_\|\,f(r_\|)$.
Note that this total cross section~\eqref{total} contains all three Bragg peaks, 
the main peak centered at ${\bf\Delta k}=0$ as well as the two side peaks centered at 
$\Delta{\bf k}=\pm2{\bf k}_1$.
This can be understood by Fourier decomposition of the standing wave profile 
$j^{\rm eff}\propto\cos^2({\bf k}_1\cdot{\bf r})=[1+\cos(2{\bf k}_1\cdot{\bf r})]/2$.
Hence, each side peak has roughly half the amplitude of the main peak at ${\bf\Delta k}=0$. 

Focusing on the experimentally most relevant side peaks at $\Delta{\bf k}=\pm2{\bf k}_1$, 
we find the total cross section for the case ${\bf e}_{\rm in}={\bf e}_{\rm out}={\bf e}_1$
after spatial and temporal average to be $\sigma_\pm\approx12\omega^2_{\rm out}L_\|^2A_\perp a^2 E_1^4$
%
%
where $A_\perp$ is the effective focus area seen by the XFEL. 
Roughly speaking, the ratio $\sigma_\pm/A_\perp$ determines the probability that an XFEL photon 
hitting the optical laser focus will get scattered. 
Inserting the QED value for $a$ we find 
\bea
\label{total-cross-section}
\sigma_\pm\approx12A_\perp
\left(\omega_{\rm out}L_\|\,\frac{\alpha_{\rm QED}}{90\pi}\,\frac{E_1^2}{E_{\rm crit}^2}\right)^2  
\,,
\ea
where $\alpha_{\rm QED}\approx1/137$ is the fine structure constant. 

For the other XFEL polarization ${\bf e}_{\rm in}={\bf e}_{\rm out}=\pm{\bf n}_1$, 
the magnetic field of the XFEL interacts with the electric field of the optical laser via the 
$b$ term in Eq.~\eqref{EHL}.  
Inserting the QED prediction $b=7a$, the signal would be a factor of $(7/4)^2\approx3$ higher,
but otherwise the same conclusions as above apply. 
Furthermore, this channel would also be sensitive to potential axion or axion-like particles.

\section{``Five O'Clock'' Scenario}\label{``Five O'Clock'' Scenario}

As 
a scenario where the final x-ray photon does also receive an energy shift,
we consider the superposition of two optical lasers with different frequencies, such as 
$\omega_2=2\omega_1$ (which could be generated by frequency doubling, for example). 
Keeping the XFEL perpendicular to the first optical laser ${\bf n}_{\rm in}\perp{\bf n}_1$, 
we may satisfy energy and momentum conservation by tilting the second laser by 30 degrees such that 
${\bf n}_{\rm in}\cdot{\bf n}_2=1/2$. 
The resulting momentum transfer in forward direction is then consistent with the energy shift 
$\omega_{\rm out}=\omega_{\rm in}\pm\omega_1$. 
To maximize the transversal momentum transfer (i.e., the scattering angle), we may choose an 
orientation where ${\bf n}_2$ lies in the same plane as ${\bf n}_1$ and ${\bf n}_{\rm in}$
but almost opposite to ${\bf n}_1$, i.e., ${\bf n}_1\cdot{\bf n}_2=-\sqrt{3}/2$. 
Picturing ${\bf n}_1$ as vertical and ${\bf n}_{\rm in}$ as horizontal, ${\bf n}_1$ and ${\bf n}_2$ 
look like the hands of a clock at five or seven o'clock, see Fig.~\ref{figure}c.

Naturally, the angular dependence~\eqref{general-F} of ${\mathfrak F}$ 
is a bit more involved than in the previous sections, 
but the general behavior is quite similar.
For example, if all polarizations ${\bf e}_{\rm in}={\bf e}_{\rm out}={\bf e}_1={\bf e}_2$ 
are perpendicular to the plane spanned by ${\bf n}_1$ and ${\bf n}_2$, 
we find that only the $a$-term contributes ${\mathfrak F}=2a$, analogous to Eq.~\eqref{crossed}. 
As before, there is no polarization flip in this specific and highly symmetric case 
${\bf n}_1\perp{\bf e}_{\rm in}={\bf e}_{\rm out}={\bf e}_1={\bf e}_2\perp{\bf n}_2$,
but for most other orientations, we do also obtain a birefringent signal.  

Note that the ``five o'clock'' scenario considered here is different from the ``y-scenario'' 
studied in~\cite{king-18}. 
Although both feature the same energy shift $\omega_{\rm out}=\omega_{\rm in}\pm\omega_1$, 
the ``five o'clock'' scenario offers a larger momentum transfer and thus scattering angle.

\section{Experimental Parameters}

Let us discuss possible experimental realizations and estimate the order of magnitude of the expected signal, 
where we take the experimental capabilities at the Helmholtz International Beamline for Extreme Fields (HIBEF) 
as an example. 
As a High Energy Density (HED) instrument \cite{ULF}, the European XFEL is combined with the Relax 
laser \cite{relax} provided by the HIBEF user consortium. 
%
%
We start with a conservative estimate and insert values which have already been shown to be reachable 
experimentally.
The optical laser is characterized by its frequency $\omega_{\rm L}=1.5~\rm eV$,  
focus intensity $3.5\times10^{20}\rm W/cm^2$, as well as focus width $2.5~\mu\rm m$ and 
length $9~\mu\rm m$ (corresponding to a duration of 30~fs) \cite{relax}. 

For the XFEL, we assume a frequency $\omega_{\rm in}=6\,\rm keV$ with $10^{12}$ photons per pulse 
(corresponding to an energy of 1~mJ), focused to a width between 4 and 5~$\mu\rm m$ 
with a beam divergence of $80~\mu\rm rad$ \cite{ULF}. 
A tighter focus is possible with a different lensing system (e.g., the CRL4 lens \cite{ULF}), 
but then the beam divergence would increase by an order of magnitude. 


However, in view of the scattering angle of $500~\mu\rm rad$ 
(for $\omega_{\rm L}=1.5~\rm eV$ and $\omega_{\rm in}=6~\rm keV$) for the crossed-beam scenario,  
it is probably better to employ the lower beam divergence 
(or use more involved schemes, such as the dark-field method, see, e.g., \cite{karb-22}). 
Placing the detector at a distance of 7~meters to the interaction point, the scattering angle of 
$500~\mu\rm rad$ yields a deflection by 3.5~mm which should by sufficient to distinguish the 
scattered signal from the XFEL beam. 
%


Now we are in the position to provide a rough estimate of the number of scattered photons in such 
an experiment. 
As already explained in Sec.~\ref{Cross Section}, the finite size of the optical laser focus 
leads to a cut-off for the spatial integral in the amplitude~\eqref{amplitude} in terms of 
the effective focus volume $V_{\rm eff}$. 
%
%
Thus, the differential cross section scales as 
(up to dimensionless kinematical factors like $\mathfrak{F}_a$ and $\mathfrak{F}_b$
as well as spatial and temporal overlap integrals)
%
%
\bea
\frac{d\sigma}{d\Omega}
=
\ord\left(
\frac{\alpha_{\rm QED}^2}{(360\pi^2)^2}\, 
\frac{E_{\rm L}^4}{E_{\rm crit}^4}\,
\omega_{\rm in}^4V_{\rm eff}^2\right)
\,.
\ea
Inserting the values above, this differential cross section becomes 
$d\sigma/d\Omega=\ord(10^{-8}\mu\rm m^2)$. 
However, as also explained in Sec.~\ref{Cross Section}, this value is only valid in a comparably small 
solid angle of $\Delta\Omega=\ord(10^{-9})$ corresponding to the size of the diffraction peak, 
which is determined by the spatial extent of the optical laser focus. 
%
%
Thus, the total cross section for this Bragg peak reads $\sigma=\ord(10^{-17}\mu\rm m^2)$ 
which corresponds to Eq.~\eqref{total-cross-section}. 

As a result, one obtains $\ord(10^{-6})$ signal photons per shot, or one signal photon per $\ord(10^{6})$ shots.
Even with a repetition rate of 5~Hz, $\ord(10^{6})$ shots correspond to 
one day of continuous measurements.
%
%
While not impossible, such an experiment would certainly be extremely challenging. 
The smallness of the signal again demonstrates the paramount importance of suppressing the background 
as much as possible.

Thus, let us discuss options to enhance the signal.
The most obvious possibility is to increase the optical laser intensity since the signal scales with the 
square of that quantity. 
An intensity of $10^{21}\rm W/cm^2$ 
is already technically available and shall be provided in near future.
This would enhance the signal by one order of magnitude. 
%
%
%
Further upgrades should enable us to reach $10^{22}\rm W/cm^2$ yielding another increase of the signal 
by two additional orders of magnitude. 
%

In principle, increasing the XFEL frequency (to 12~keV, for example) also enhances the cross section, 
but, on the other hand, reduces the scattering angle (if the optical laser frequency is kept constant) 
and typically lowers the number of XFEL photons.
Thus, this is perhaps not the best option for improvement. 
Of course, increasing the number of XFEL photons (with all other relevant quantities staying the same) 
would be advantageous.  
Similarly, a larger volume $V_{\rm eff}$ of the optical laser focus, 
as long as it is not at the expense of the intensity, would increase the scattering yield.

%
%

One way to realize the collision of the two optical pulses could be the set-up already proposed in 
\cite{heinzl-06}, for example, where each laser pulse, after the collision at the focus, hits the 
parabolic mirror used to focus the other laser pulse -- and thus retraces its optical path. 
In this scenario, the two laser pulses are basically the time reversals of each other, which 
requires some fine-tuning of the optical paths. 
A way to avoid this fine-tuning and potential damage is to tilt both optical axes a bit 
(corresponding to a ``five-past-five'' geometry, see Fig.~\ref{figure}d) 
such that the optical paths of the two pulses only overlap at the focus. 
This ``five-past-five'' geometry would reduce the momentum transfer and thus scattering angle a bit, 
but might be easier to realize experimentally. 

Developing this idea further, one could also imagine tilting the optical axes even more, e.g., 
in the form of a ``ten-past-four'' geometry, see Fig.~\ref{figure}e. 
In this way, one could interpolate between the crossed-beam case in Sec.~\ref{Crossed Beam Case}
and the counter-propagating scenario in Sec.~\ref{Counter-propagating Case}. 
Going from the crossed-beam to the counter-propagating case has two main advantages.
First, the amplitude increases, compare Eqs.~\eqref{counter-propagating} and \eqref{crossed} and the Appendix. 
Second, the interacting length is enlarged. 
Both would enhance the signal strength. 
As a drawback, the momentum transfer and thus scattering angle is reduced.
Hence, an optimum tilt angle is determined by the trade-off between signal strength and background 
suppression (as well as experimental constraints). 

As in many of the other proposals for detecting vacuum birefringence, the polarization of the x-ray 
photons can be distinguished via Bragg reflection crystals. 
Those can be designed to also provide the energy resolution which would allow us to detect the energy 
shift in the ``five-o'clock'' scenario.
%
%
%
%
The required frequency doubling of one of the laser pulses could be achieved with nonlinear crystals.
So far, this process has been demonstrated with an efficiency above 10~\%, 
which should be increased in the future.   
Having achieved a sufficient efficiency (or initial laser power), 
one could also imagine realizing the crossed-beam scenario 
with two frequency doubled pulses, which would imply doubling the momentum transfer and thus scattering angle.

\section{Conclusions}

%

\begin{figure}[h]
    \centering
    \includegraphics[width=0.5\textwidth]{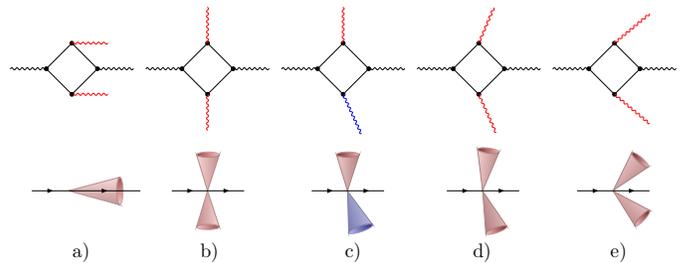}
\caption{Sketch of the counter-propagating (a), crossed-beam (b), ``five-o'clock'' (c), 
``five-past-five'' (d), and ``ten-past-four'' (e) scenarios (from left to right). 
In the bottom row, the XFEL photons are depicted as horizontal black lines while the focused optical lasers  
are represented by red ($\omega=1.5~\rm eV$) or blue ($\omega=3~\rm eV$) cones. 
The top row displays a typical Feynman diagram where the color coding and the angle idicate which beam the 
involved photon lines belong to.}
\label{figure} 
\end{figure}

As an example of light-by-light scattering, we study the interaction of x-ray photons with ultra-strong optical
lasers and compare different scenarios.
Apart from the the counter-propagating case already discussed in the literature, we consider 
the crossed-beam case, the ``five-o'clock'' scenario, as well as interpolating cases such as the 
``five-past-five'' and the ``ten-past-four'' scenarios, see Fig.~\ref{figure}. 
All cases yield scattering amplitudes of comparable order of magnitude, facilitate birefringent
scattering, and allow us to address the $a$ and $b$ parameters in the effective Euler-Heisenberg
Lagrangian separately via adjusting the polarization vectors accordingly.

As a difference, the interaction length is set by the pulse length of the optical laser focus in the first
(counter-propagating) case while it is mainly determined by the focal width in the 
crossed-beam case (and accordingly for the other scenarios). 
More important, however, is the distinction between the initial and final x-ray photons, which allows us
to discriminate them from the background.
In the first (counter-propagating) case, the only measurable difference is their polarization --
at least in the plane-wave approximation.
A finite scattering angle (corresponding to a non-zero momentum transfer) can only be induced by the
spatial inhomogeneity of the laser focus, which makes it comparably small effect, see also \cite{gies-15}.
In contrast, the other two scenarios lead to a significantly larger momentum transfer.
For example, taking an optical laser with $\omega_1=1.5~\rm eV$ and an XFEL with 6~keV,   
we find scattering angles of around half a mrad,  
which helps us to separate the signal photons from the background (i.e., the main XFEL beam).

Furthermore, the  
``five-o'clock'' scenario -- while a bit more challenging to set up experimentally --
would also yield an energy shift (of 1.5~eV in our example) of the final photons, 
which provides yet another important
channel for separating signal and background.
These findings motivate further studies and give rise to the hope for observing this fundamental QED
phenomenon at experimental facilities such as HIBEF.






\acknowledgments 
The authors acknowledge fruitful discussions with H.~Gies, F.~Karbstein, and R.~Sauerbrey 
as well as funding by the Deutsche Forschungsgemeinschaft 
(DFG, German Research Foundation) -- Project-ID 278162697-- SFB 1242.

\appendix 


\section{Planar Configuration}
In this section we present a formula for the most general planar case where all the wave vectors (momenta) 
of the pump as well as the probe laser lie on a plane, 
which for simplicity can be taken to be the $xy$-plane. 
We further express
\bea
&&{\bf n}_1\rightarrow \cos\theta_1{\bf e}_x 
+\sin\theta_1{\bf e}_y
\;,\, 
{\bf n}_{\rm in}\rightarrow{\bf e}_y 
\,,
\nonumber\\
&&{\bf n}_2\rightarrow \cos\theta_2{\bf e}_x 
+\sin\theta_2{\bf e}_y
\;,\, 
\omega_2=l\omega_1
\,,
\ea
where $\theta_i$ is the angle that the unit vector ${\bf n}_i$ makes with the $x$ axis, 
while $l>0$ denotes the ratio of the two optical laser frequencies.
%
%
Up to order $\mathcal{O}(\omega_{\rm in}^{-1})$, the following relation is thus to be satisfied if one desires to stay in the kinematically allowed region 
\bea
\sin\theta_2=\frac{\sin\theta_1+l-1}{l}.
\label{theta2}
\ea
Choosing the following parametrization of the laser's and XFEL's polarizations 
\bea
{\bf e}_1&=&(\sin\alpha_1\sin\theta_1,-\sin\alpha_1\cos\theta_1,\cos\alpha_1),\nonumber\\
{\bf e}_2&=&(-\sin\alpha_2\sin\theta_2,\sin\alpha_2\cos\theta_2,\cos\alpha_2),\nonumber\\
{\bf e}_{\rm in}&=&(\sin\alpha_{\rm in},0,\cos\alpha_{\rm in}),\nonumber\\
{\bf e}_{\rm out}&=&(\sin\alpha_{\rm out},0,\cos\alpha_{\rm out}),
\ea
where $\alpha_{i}$ $(i=1,2,{\rm in},{\rm out})$ is the angle formed by the polarization vector $\mathbf{e}_{i}$ with respect to $\hat{z}$,
the function $\mathfrak{F}$ in this scenario is finally given by 
\bea
\mathfrak{F}&=&\frac{1}{2}(1-\sin\theta_1)(1-\sin\theta_2)\nonumber\\
&&\times\Big[(4a+b)\cos(\alpha_1+\alpha_2)\cos(\alpha_{\rm in}-\alpha_{\rm out})\nonumber\\
&&\hspace{0.3cm}+(4a-b)\cos(\alpha_1-\alpha_2+\alpha_{\rm in}+\alpha_{\rm out})\Big].
\ea
Recall that $\theta_2$ is fixed by Eq.~(\ref{theta2}) if one desires to obtain a nonvanishing amplitude.

\end{document}